\documentclass[b5paper,twoside,reqno]{bjp}
\usepackage{cite}
\usepackage{latexsym}
\usepackage{graphicx}
\usepackage{amssymb,amsmath,amsfonts,amsbsy, amsthm, xspace, latexsym, amscd, enumitem}
\usepackage[utf8]{inputenc}
\usepackage[T1]{fontenc}
\usepackage{epstopdf}
\usepackage[english]{babel}
\usepackage{times}
\usepackage{url,bm}
\renewcommand\congress{\small\sf Section: Theoretical Physics}

\makeindex

\begin{document}

\sloppy \raggedbottom

\title{Electrostatics of quadrupolarizable media}

\runningheads{Electrostatics of quadrupolarizable media}{Slavchov, Dimitrova, and Ivanov}

\begin{start}
\coauthor{Radomir I. Slavchov}{1,2}, \author{Iglika M. Dimitrova}{1}, \\ \coauthor{Tzanko I. Ivanov}{2}

\address{Department of Physical Chemistry, Faculty of Chemistry and Pharmacy, Sofia University, 1164 Sofia, Bulgaria}{1}

\address{Department of Chemical Engineering and Biotechnology, Cambridge University,UK, CB2 3RA Cambridge}{2}

\address{Department of Theoretical Physics, Faculty of Physics, Sofia University,1164 Sofia, Bulgaria}{3}

\begin{Abstract}
The classical macroscopic Maxwell equations are approximated. They are a corollary of the multipole expansion of the local electrostatic potential up to dipolar terms. But quadrupolarization of the medium should not be neglected if the molecules which build up the medium possess large quadrupole moment or do not have any dipole moment. If we include the quadrupolar terms in Maxwell equations we obtain the quadrupolar analogue of Poisson's equation: $\nabla^2 \phi  - L^2_Q\nabla^4 \phi = - \rho / \varepsilon$. This equation is of the fourth order and it requires not only the two classical boundary conditions but also two additional ones: continuous electric field and the relation of the jump of the normal quadrupolarizability at the surface to the intrinsic normal surface dipole moment. The account of the quadrupole moment of the molecules leads to significant differences compared to the classical electrostatic theory.
\end{Abstract}


\end{start}

\section[]{Introduction}\label{Sec. 1}
The macroscopic Coulomb and Ampere's law are~\cite{ref01}:
\begin{equation}\label{Eq1}
    \nabla\cdot\bf {D}=\rho
\end{equation}
\begin{equation}\label{Eq2}
    {\bf E}=-\nabla\phi.
\end{equation}
Here,  $\rho$ is the free charge density,  $\phi$ is the electrostatic potential and $\bf {D}$ is the electric displacement field which is linearly dependent on the electric field intensity ${\bf E}$ \cite{ref02}:
\begin{equation}\label{Eq3}
\bf {D}\equiv \varepsilon_0 \bf {E}+\bf {P}=\varepsilon_0 \bf {E}+\alpha_P\bf {E}=\varepsilon \bf {E}
\end{equation}
where $\varepsilon=\varepsilon_0 +\alpha_P =  \varepsilon_0 \varepsilon_\mathrm{r}$ is the absolute dielectric permittivity, $\varepsilon_0$  is the vacuum permittivity, $\varepsilon_\mathrm{r}$  is the relative permittivity of the medium,  $\alpha_P$ is the macroscopic polarizability of the medium. For a homogeneous medium ($\nabla\varepsilon   = 0$) the Poisson equation for $\phi$  follows from Eqs.\eqref{Eq1}-\eqref{Eq3}:
\begin{equation}\label{Eq4}
    -\varepsilon \nabla^2 \phi=\rho.
\end{equation}
The derivation of Eq.\eqref{Eq4} involves a multipole expansion of the local potential up to dipole terms, i.e., it neglects the quadrupole moment density \cite{ref01, ref03,ref04}.  Several studies of optical phenomena \cite{ref05, ref06,ref07,ref08} have demonstrated that the quadrupolar terms in the macroscopic Coulomb law Eq. \eqref{Eq1} become quite significant in cases where high gradients of ${\bf E}$ are present. In such cases, quadrupolar term in the displacement field ${\bf D}$ need to be introduced \cite{ref08}:
\begin{equation}\label{Eq5}
    \bf {D}= \varepsilon_0 \bf {E}+\bf {P}-\frac{1}{2}\nabla\cdot{\bf Q}.
\end{equation}
Here,${\bf Q}$ is the macroscopic density of the quadrupole moment tensor (with zero trace \cite{ref09}). Note that the numerical coefficient in front of $\nabla\cdot{\bf Q}$ depends on the choice of definition of the microscopic quadrupole moment ${\bf q}$ - we used the following  one \cite{ref09}:
\begin{equation}\label{Eq6}
 {\bf q}=\int\limits_{\mathrm{particle}} \left [{\bf r}{\bf r}-\frac{1}{3}{\bf U}r^2 \right ] \rho_{\mathrm{local}}({\bf r}) \mathrm{d}V,
\end{equation}
where  $\rho_{\mathrm{local}}$ is the local (microscopic) charge density in the particle and ${\bf U}$ is the unit tensor. Other definitions of ${\bf q}$ are often employed, differing from Eq. \eqref{Eq6} with a factor of 3 \cite{ref01} or 3/2 \cite{ref04}.

In order to close the problem, in addition to Eqs. \eqref{Eq1} and \eqref{Eq5}, one needs a constitutive relation between quadrupole moment density and electric field gradient. The equation of state of the quadrupolarization is \cite{ref09}:
\begin{equation}\label{Eq7}
    {\bf Q}=\alpha_Q\left (\nabla{\bf E}-\frac{1}{3}{\bf U}\nabla\cdot{\bf E} \right ).
\end{equation}
Since it is of key importance for the theory of quadrupolar dielectrics, we will present shortly its derivation in Sec. \ref{Sec. 2}. Here, the coefficient $\alpha_Q$ is the quadrupolarizability of the medium and it can be related to the quadrupole moment of the solvent molecules \cite{ref04,ref09}. Various other constitutive relations have been proposed in the literature \cite{ref03,ref04}. Substituting Eqs. \eqref{Eq5}, \eqref{Eq7} into Eq.\eqref{Eq1} and using the Ampere's law (${\bf E}=-\nabla\phi$, Eq. \eqref{Eq2}) and the equation of state of the polarization $\bf {P} =  \alpha_P \bf {E}$, one obtains the explicit form of the electrostatic Coulomb-Ampere's law in quadrupolarizable media:
\begin{equation}\label{Eq8}
   \nabla^2 \phi  - L^2_Q\nabla^4 \phi = - \rho / \varepsilon
\end{equation}
which determines the electrostatic potential $\phi$. Here, the quadrupolar length $L_Q$ is defined as:
\begin{equation}\label{Eq9}
   L^2_Q=\frac{\alpha_Q}{3\varepsilon}.
\end{equation}
In Ref. \cite{ref09}, we used data for the partial molar volumes and entropies for various ions to estimate this quantity for water: $L_Q = 2.5 \pm 1.5$ {\AA}. Equations \eqref{Eq8} and \eqref{Eq9} are of the same form as those of Chitanvis \cite{ref03}, the only difference being the obtained different numerical coefficient in Eq. \eqref{Eq9}. At $L_Q = 0$, the quadrupolar Coulomb-Ampere's law Eq. \eqref{Eq8} simplifies to the standard Poisson equation (Eq. \eqref{Eq4}).
Equation \eqref{Eq8} opens a vast field for analysis of the effect of the quadrupole moments of the molecules composing a medium on many electrostatic phenomena. The correction for ${\bf Q}$ will be important if the solvent molecules possess large quadrupole moment - such is the case of water \cite{ref10} and many others, including "non-polar" media of low dipole moment but high quadrupole moment such as liquid CO$_2$, fluorocarbons, etc. \cite{ref04,ref11}.

\section[]{	Equation of state for the quadrupole moment density}\label{Sec. 2}

The problem for the constitutive relation between ${\bf Q}$ and the field gradient $\nabla{\bf E}$ has been addressed several times \cite{ref03, ref04, ref05, ref08, ref12, ref13, ref14, ref15}. Using as a starting point the approach of Jeon and Kim \cite{ref04}, we obtained in Ref. \cite{ref09} a new simple equation of state which relates ${\bf Q}$ to the field gradient $\nabla{\bf E}$ and the molecular properties of the solvent (Eq. \eqref{Eq7}). Here we will mark the basic points of that derivation.

Consider an ideal gas consisting of molecules possessing a solid quadrupole moment tensor ${\bf q_0}$ (for the sake of simplicity, the molecule is assumed non-polarizable and with no dipole moment). Since ${\bf q_0}$ is symmetrical and traceless, by a suitable choice of the coordinate system it can be diagonalized \cite{ref05} and in the general case, its diagonal form is:
\begin{equation}\label{Eq10}
 {\bf q_0} = \left(
               \begin{array}{ccc}
                 q_{xx} & 0 & 0 \\
                 0 & q_{yy} & 0 \\
                 0 & 0 & q_{zz} \\
               \end{array}
             \right) - \frac{q_{xx}+q_{yy}+q_{zz}}{3}\bf {U}.
\end{equation}
Here we remove the trace of the quadrupole moment because it causes a constant potential (Bethe potential \cite{ref16}) which has no contribution to the electric field \cite{ref08}. The molecule is freely rotating. In a rotated frame the quadrupole moment tensor changes from ${\bf q_0}$ to ${\bf q}$:
\begin{equation}\label{Eq11}
    q_{ij}(\varphi,\psi,\theta)= E_{ik}(\varphi,\psi,\theta)E_{jl}(\varphi,\psi,\theta)q_{0kl}
\end{equation}
where $\varphi,\psi$ and $\theta$ are the Eulerian angles and $\textrm{E}(\varphi,\psi,\theta)$ is the Euler matrix.
In the absence of a gradient of the electric field the average value of ${\bf q}$ is $\bf{q_0}$. In an external electric field gradient $\nabla{\bf E}$, the electric energy of the molecule is given by the expression ({\it Eq 4.17} of Jackson \cite{ref01}):
\begin{equation}\label{Eq12}
    u_{\mathrm{el}}=-\frac{1}{2}{\bf q}:\nabla{\bf E}.
\end{equation}
The symbol “:” denotes double scalar product, ${\bf A}:{\bf B} = A_{ij}B_{ji}$.The probability for a given orientation of the molecule follows the Boltzmann distribution which can be linearized in the case of $u_{\mathrm{el}}/k_\mathrm{B}T \ll 1$:
\begin{equation}\label{Eq13}
\rho(\varphi,\psi,\theta)=c_\mathrm{n} \exp\left({-\frac{u_{\mathrm{el}}}{k_\mathrm{B}T}}\right) \approx c_\mathrm{n} \left (1-\frac{u_{\mathrm{el}}}{k_\mathrm{B}T} \right ).
\end{equation}
Here, $k_\mathrm{B}$ is the Boltzmann constant, $T$ is the absolute temperature and $c_\mathrm{n}$ is a normalizing coefficient which can be obtained from the condition $\int \rho(\varphi,\psi,\theta) \mathrm{d}\Omega=1$.
The average quadrupole moment $\bar{\bf {q}}$  of a molecule can be calculated directly using Eqs. \eqref{Eq10}-\eqref{Eq13}:
\begin{equation}\label{Eq14}	
\bar{\bf {q}}=\int\limits^{2\pi}_0 \int\limits^{2\pi}_0 \int\limits^{\pi}_0 \bf {q} \rho(\varphi,\psi,\theta) \sin\theta d \theta d \varphi d \psi=\alpha_q\left (\nabla{\bf E}-\frac{1}{3}{\bf U}\nabla\cdot{\bf E} \right )
\end{equation}
Here, we have introduced the molecular quadrupolarizability $\alpha_q$ which is related to the diagonal components of ${\bf q_0}$ as follows:
\begin{equation}\label{Eq15}
\alpha_q={\bf q_0}:{\bf q_0}/10k_\mathrm{B}T
\end{equation}
Equation \eqref{Eq15} was obtained e.g. in Ref. \cite{ref04}. The derivation above is strictly valid for a gas of solid quadrupoles. It can be readily generalized to include molecular quadrupolarizabilities, $\alpha_{q0}$ \cite{ref04} and then we obtain the expression:
\begin{equation}\label{Eq16}
\alpha_q=\alpha_{q0}+{\bf q_0}:{\bf q_0}/10k_\mathrm{B}T
\end{equation}
	In the presence of a field gradient $\nabla{\bf  E}$, the macroscopic density ${\bf Q}$ of the quadrupole moment in a gas is the gas concentration $C$ times $\bar{\bf {q}}$, Eq. \eqref{Eq14}. Therefore, we finally obtain Eq. \eqref{Eq7} with macroscopic quadrupolarizability defined as $\alpha_Q = C \alpha_q$. The relation $\alpha_Q \sim C(\alpha_{q0}+{\bf q_0}:{\bf q_0}/10k_\mathrm{B}T)$ can be compared to the linear Langevin-Debye formula $\alpha_P \sim C(\alpha_{p0}+{\bf p}\cdot{\bf p}/3k_\mathrm{B}T)$ \cite{ref01,ref02} ($\alpha_{p0}$ and ${\bf p}$ are the average polarizability and the dipole moment of the solvent molecule).

\section[]{	Boundary conditions for the generalized Poisson equation}\label{Sec. 3}

The quadrupolar equation for $\phi$ (Eq. \eqref{Eq8}) is of the fourth order and requires additional boundary conditions compared to Poisson’s equation. One of these new boundary conditions was deduced by Graham and Raab \cite{ref07, ref17} and by Batygin and Toptygin \cite{ref13}, and it explicitly relates {\it the intrinsic surface normal dipole moment} $P^\mathrm{S}_z$ to the bulk quadrupole densities.
Following Graham and Raab \cite{ref07}, we will derive the boundary conditions using the singular distribution approach developed by Albano, Bedeaux and Vlieger \cite{ref18,ref19}. We investigate a flat interface at $z = z_0$ between two quadrupolar dielectrics; this interface has surface charge density $\rho^\mathrm{S}$ and intrinsic surface dipole moment density ${\bf P}^\mathrm{S}$. First, we write the singular distributions of $\rho, \bf {P}$ and $\bf {Q}$:
\begin{equation}\label{Eq17}
\rho=\eta^+ \rho^+ + \eta^- \rho^- + \delta \rho^\mathrm{S},
\end{equation}
\begin{equation}\label{Eq18}
\bf {P}= \eta^+ \bf {P}^+ + \eta^- \bf {P}^- +\delta{\bf P}^\mathrm{S}
\end{equation}
\begin{equation}\label{Eq19}
\bf {Q}= \eta^+\bf {Q}^+ +\eta^-\bf {Q}^-.
\end{equation}
Here, $X^+$ and $X^-$ denote the corresponding physical quantities for the phase situated at $z > z_0$ and $z < z_0$, respectively; $\eta$ is the Heaviside step function, $\eta^+ \equiv \eta(z-z_0)$, $\eta^- \equiv \eta(z_0-z)$; $\delta \equiv \delta(z-z_0)$ is the Dirac delta function. If we want to include the surface excess of the quadrupole moment density, we should take into account the bulk octupole moment density. The electric field is intensive variable and so its singular distribution is:
\begin{equation}\label{Eq20}
\bf {E}= \eta^+\bf {E}^++ \eta^-\bf {E}^-.
\end{equation}
The distributions of $\bf {P}, \bf {Q}$ and $\bf {E}$ (Eqs. \eqref{Eq18}-\eqref{Eq20}) are substituted in Eq. \eqref{Eq5} to obtain the singular distribution of $\bf {D}$:
\begin{equation}\label{Eq21}
\bf {D}= \eta^+{\bf D}^+ + \eta^- \bf {D}^- +\delta {\bf D}^\mathrm{S}
\end{equation}
where $\bf {D}^+$ and $\bf {D}^-$ are the displacement fields for the upper and lower phase, respectively:
\begin{equation}\label{Eq22}
    \bf {D^{\pm}}= \varepsilon_0 \bf {E^{\pm}}+\bf {P^{\pm}}-\frac{1}{2}\nabla\cdot{\bf Q^{\pm}}.
\end{equation}
and the surface excess of the electric displacement:
\begin{equation}\label{Eq23}
{\bf D}^\mathrm{S}={\bf P}^\mathrm{S} - \frac{1}{2} \left (\bf {e}_z\cdot\bf {Q}^+-\bf {e}_z\cdot\bf {Q}^-\right).
\end{equation}
In the derivation of the last equations we used the relation $\nabla\cdot(\eta^{\pm}\bf {Q}^{\pm}) = \eta^{\pm} \nabla\cdot\bf {Q}^{\pm} \pm \delta \bf {e}_z\cdot\bf {Q}^{\pm}$ and that the Dirac delta function is a derivative of the Heaviside step function: $\nabla\eta^{\pm} = \bf {e}_z \mathrm{d} \eta^{\pm}/\mathrm{d} z = \pm\bf {e}_z\delta$. The singular distributions Eqs. \eqref{Eq21} and \eqref{Eq17} of $\bf {D}$ and $\rho$ are then substituted into Coulomb’s law Eq. \eqref{Eq1} to obtain the singular expansion of the quadrupolar Maxwell equation:	
\begin{eqnarray} \label{Eq24}
		\eta^+(\nabla\cdot\bf {D}^+ -\rho^+)+\eta^-(\nabla\cdot\bf {D}^- -\rho^-) &+& \delta(D^+_z-D^-_z+\nabla\cdot{\bf D}^\mathrm{S}-\rho^\mathrm{S}) \nonumber \\
   &+& \delta_1D^\mathrm{S}_z=0
\end{eqnarray}
where $\delta_1 = \mathrm{d}\delta/\mathrm{d} z$.  The above equation further simplifies to:
\begin{eqnarray} \label{Eq25}
		\eta^+(\nabla\cdot\bf {D}^+ -\rho^+)+\eta^-(\nabla\cdot\bf {D}^- -\rho^-) &+& \delta(D^+_z-D^-_z+\nabla^\mathrm{S}\cdot\bf {D}^\mathrm{S}-\rho^\mathrm{S})|_{z=z_0} \nonumber \\
   &+& \delta_1D^\mathrm{S}_z|_{z=z_0}=0
\end{eqnarray}
Here, we have used the properties of the singular functions: $\delta f(z) = \delta f(z_0)$ and $\delta_1 f(z) = \delta_1 f(z_0)-\delta(\mathrm{d} f/\mathrm{d} z)|_{z=z_0}$ and $\nabla^\mathrm{S}$ denotes surface tangential derivative (in flat symmetry $\nabla^\mathrm{S} = \bf {e}_x\partial/\partial x + \bf {e}_y\partial/\partial y)$. Next, we use the linear independence of $\eta^{\pm}$ and $\delta$ to decompose Eq. \eqref{Eq25} to obtain, first, the bulk equations for the two phases (the coefficients of $\eta^{\pm}$ in Eq. \eqref{Eq25}):
\begin{equation}\label{Eq26}
	\nabla\cdot\bf {D}^{\pm}=\rho^{\pm}
\end{equation}
Further, the coefficient of $\delta$ in Eq. \eqref{Eq25} has to be 0, which gives a generalization of the Gauss law for the quadrupolar media:
\begin{equation}\label{Eq27}
D^+_z-D^-_z+\nabla^\mathrm{S}\cdot{\bf D}^\mathrm{S}-\rho^\mathrm{S}=0
\end{equation}
The last term of Eq. \eqref{Eq25}, proportional to $\delta_1$, results in a new boundary condition, which relates the intrinsic surface dipole moment $P^\mathrm{S}_z$ to the jump of the quadrupole moment , cf. Eq. \eqref{Eq23}:
\begin{equation}\label{Eq28}
Q^+_{zz}-Q^-_{zz}=2P^\mathrm{S}_z
\end{equation}
This equation was derived with the classical methods by Batygin and Toptygin \cite{ref13}; compare also to {\it Eq. 65} of Shen and Hu \cite{ref20}. We will refer to it as to {\it multipolar (dipolar) condition for the jump of the electric field gradient}.  We will consider only flat symmetry in this study (the quadrupolarization tensor has diagonal elements only) and surfaces with no tangential polarization. Therefore, Eq. \eqref{Eq27} simplifies to:
\begin{equation}\label{Eq29}
D^+_z-D^-_z=\rho^\mathrm{S},
\end{equation}
which is formally equivalent to the classical Gauss law, but one must keep in mind that $\textbf{D}$ involves higher derivatives of the field $\textbf{E}$, cf. Eq. \eqref{Eq5}.
Two additional boundary conditions complete the set, namely, the potential and the electric field must be continuous at $z = z_0$,
\begin{equation}\label{Eq30}
\phi^+|_{z=z_0}=\phi^-|_{z=z_0}=\phi^\mathrm{S}, \quad \textbf{E}^+|_{z=z_0}=\textbf{E}^-|_{z=z_0}=\textbf{E}^\mathrm{S}
\end{equation}
Instead of continuous $\textbf{E}$, Chitanvis imposed continuity of the second normal derivative of the normal field but the field itself remained discontinuous in his work.
Equations \eqref{Eq8} and \eqref{Eq28}-\eqref{Eq30} define a unique solution for the electrostatic potential $\phi$. Some simple consequences of it were investigated in Ref. \cite{ref09,ref21,ref22,ref23} and are summarized in the next few sections. Compared to the results of the classical dipolar electrostatics, two common features of the solutions of the quadrupolar electrostatic law are the regularization of the potential and the damping of the field gradient.

\section[]{Effect of the quadrupolarizability of the media}\label{Sec. 4}

In this Section we will apply the general equation Eq. \eqref{Eq8} of the electrostatics of quadrupolar media and its boundary conditions Eqs. \eqref{Eq28}-\eqref{Eq30} to solve several basic electrostatic problems for point sources in both isolators and conductors.

\subsection[]{Point sources in an insulator}\label{Sec. 4.1}

\begin{figure}[h]
\includegraphics[width=155pt]{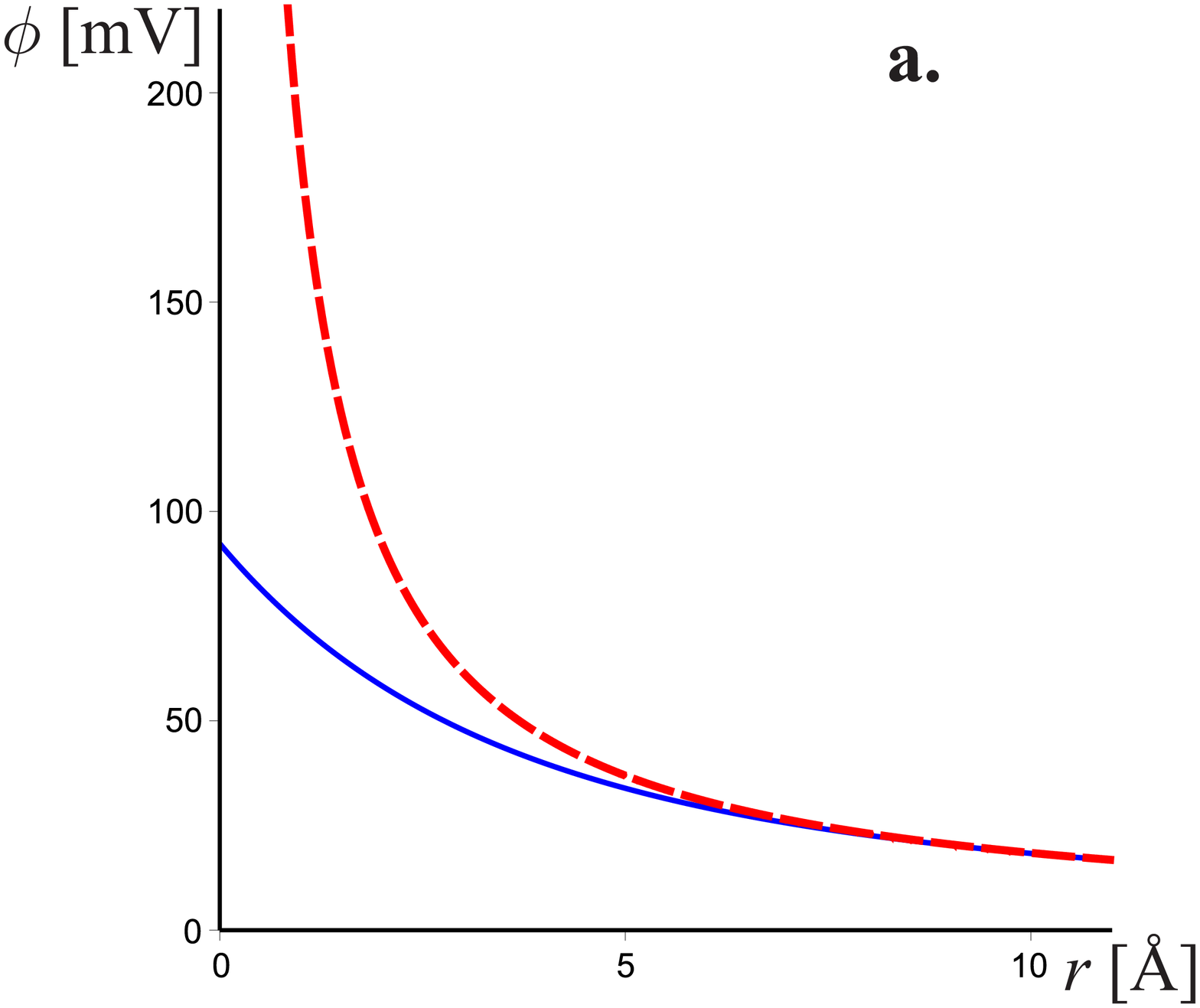}
\includegraphics[width=155pt]{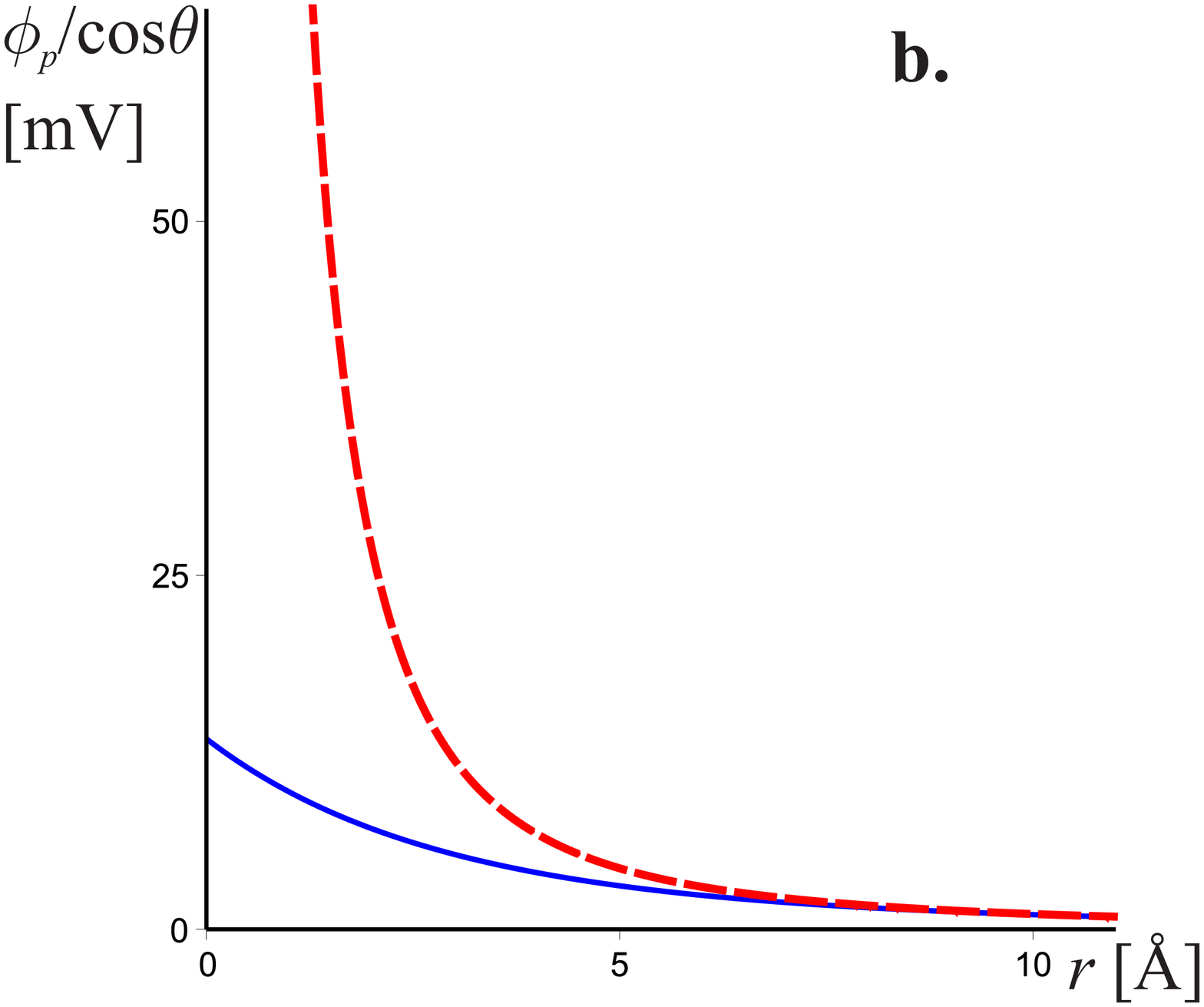}
\caption{Electrostatic potential of \textbf{a}) a point charge, Eq. \eqref{Eq33}, and \textbf{b}) a point
dipole ($|\textbf{p}| = 2.8 \mathrm{D} =  9.34\times10^{-30}$ C m), Eq. \eqref{Eq35}, in a
quadrupolar medium vs. the distance $r$ from the point charge/dipole in water
($\varepsilon = 78\times\varepsilon_0$, $L_Q = 2$ \AA). In a quadrupolar medium, the point charge has
finite potential at $r = 0$; the potential of the point dipole is also finite but
discontinuous at $r = 0$. Blue solid line: $L_Q = 2$ \AA; red dashed line: $L_Q = 0$
(the classical solution).}\label{fig:1}

\end{figure}

First, we consider a point charge with $ \rho(r)= q \delta(r)$. In this case, Eq. \eqref{Eq8} reads as:
\begin{equation}\label{Eq31}
  \frac{1}{r^2}\frac{d}{dr}r^2\frac{d}{dr}\phi-\frac{L^2_Q}{r^2}\frac{d}{dr}r^2\frac{d}{dr}\frac{1}{r^2}\frac{d}{dr}r^2\frac{d}{dr}\phi=-\frac{q \delta(r)}{\varepsilon}.
\end{equation}
The general solution of this equation is:
\begin{equation}\label{Eq32}
 \phi=A_0+\frac{A_1}{r}+A_2\frac{\mathrm{e}^{-r/L_Q}}{r}+A_3\frac{\mathrm{e}^{r/L_Q}}{r}.
\end{equation}
In order to determine the four integration constants we need to impose conditions on $\phi$. First, we require the potential to be non-divergent as $r \to \infty$ (this gives $A_3 = 0$, $A_0$ has no physical meaning and we set it to be 0). The second condition is that the asymptotic behavior of $\phi$ as $r \to \infty$ is unaffected by the presence of quadrupoles, that is, the potential of a point charge at $r \to \infty$ tends to $q/4\pi\varepsilon r$. This condition yields $A_1 = q/4\pi\varepsilon$ (note that the same result can be obtained by the Gauss law as well). We need one final condition in order to determine $A_2$. We impose the requirement that the electric field $\textbf{E}$ tends to something finite as $r \to 0$, i.e., there is no singularity of $\textbf{E}$ at $r \to 0$, which gives $A_2 = -A_1$. Thus, we obtain solution for the potential which is also finite:
\begin{equation}\label{Eq33}
  \phi=\frac{q}{4\pi\varepsilon}\frac{1-\mathrm{e}^{-r/L_Q}}{r}.
\end{equation}
The value of the potential at $r = 0$ is  $\phi_0 = q/4\pi\varepsilon L_Q$. The point charge has, therefore, a finite self-energy:
\begin{equation}\label{Eq34}
u_{\mathrm{el}}=\frac{q\phi_0}{2}=\frac{q^2}{8\pi\varepsilon L_Q}.
\end{equation}
This result is in marked contrast to the case of a point charge in vacuum where the potential is diverging as 1/r and the electrostatic self-energy of a point charge is infinite (Fig. \ref{fig:1}a). For a point charge in water at $T = 298 K$, if $L_Q = 2$ {\AA}, we obtain  $\phi_0 = 92 mV$ and $u_{\mathrm{el}} = 3.6 k_\mathrm{B}T$.  Equation 2.8 of Chitanvis \cite{ref03} has the same form as Eq. \eqref{Eq33} (but his relation between $L_Q$  and  $\alpha_Q$ is different). Equation \eqref{Eq33} can be compared also to Eq. 2.48 of Jeon and Kim \cite{ref04}, who obtained a divergent  potential since they used another constitutive relation for $\mathbf Q$ and implied different conditions on their solutions to determine the integration constants.

The potential of {\it a point dipole in quadrupolar medium} can be obtained from the point charge formula Eq. \eqref{Eq33} using the general relation  $\phi_p = -\textbf{p}\cdot \nabla \phi/q$ ($\textbf{p}$ is the dipole moment). The result is:
\begin{equation}\label{Eq35}
  \phi_p=\frac{\textbf{p}\cdot\textbf{r}}{4\pi\varepsilon r^3} \left [1-\left (1+\frac{r}{L_Q}\right )\mathrm{e}^{-r/L_Q} \right ].
\end{equation}
This potential is finite, but not continuous. It is illustrated in Fig. \ref{fig:1}b.
Finally, a point quadrupole with quadrupole moment $\textbf{q}$ in a quadrupole medium has a potential given by:
\begin{equation}\label{Eq36}
 \phi_q=\frac{3\textbf{r}\cdot\textbf{q}\cdot\textbf{r}}{8\pi\varepsilon r^5}\left [1-\frac{1}{3}\left (1+3\frac{r}{L_Q}+\frac{r^2}{L^2_Q}\right ) \mathrm{e}^{-r/L_Q} \right ].
\end{equation}
Here, we have employed the well-known formula $\phi_q=\textbf{q}:\nabla\nabla\phi/q$.

Let us summarize the results that we obtained for a point source in a quadrupolar medium. The potential of a point charge in a dipolar medium has $1/r$ singularity at $r = 0$, while it is finite and continuous in quadrupolar medium, Eq. \eqref{Eq33}. The point dipole classically has a   $\sim 1/r^2$ singularity in dipolar medium, while in a quadrupolar one it has finite (but discontinuous) potential, Eq. \eqref{Eq33}. Finally, the potential of a point quadrupole has a $1/r^3$ singularity in a dipolar and $1/r$ singularity in a quadrupolar medium, Eq. \eqref{Eq36}. It is easy to predict that in an octupolarizable medium, not only the potential, but also the field of a point dipole will be finite and continuous, and thus the self-energy  -$\textbf{p}\cdot\textbf{E}(0)$ of a dipole in an octupolar medium must be finite. A point quadrupole in octupolar medium will have a finite continuous potential, but singular $\nabla\textbf{E}$ and infinite self-energy; higher-order macroscopic multipolarizability will lead to additional regularization.

\subsection[]{Point charges in conducting media}\label{Sec. 4.2}

\begin{figure}[h]
\begin{center}
\includegraphics[width=175pt]{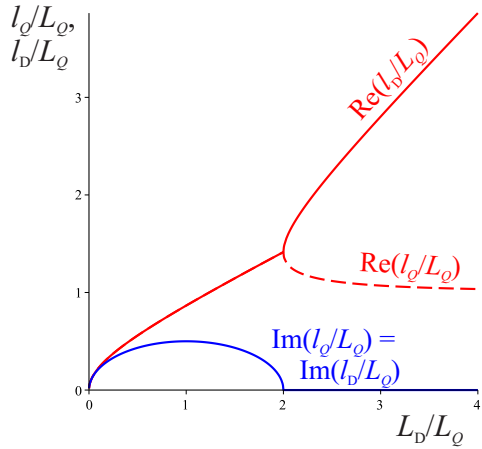}
\caption{Dimensionless characteristic lengths $l_\mathrm{D}/L_Q$ and $l_Q/L_Q$ as functions of the dimensionless Debye length $L_\mathrm{D}/L_Q$, Eq. \eqref{Eq42}. Red solid line: $\mathrm{Re}(l_\mathrm{D}/L_Q)$; red dashed line: $\mathrm{Re}(l_\mathrm{D}/L_Q)$; blue lines: $\mathrm{Im}(l_Q/L_Q)$ and $\mathrm{Im}(l_\mathrm{D}/L_Q)$.}\label{fig:2}
\end{center}
\end{figure}

In the case of conducting media, one has to consider the charge density of the mobile charges $\rho_{\mathrm{mobile}}$. We need to know the dependence of $\rho_{\mathrm{mobile}}$ on the electrostatic potential. In this work, we assume that the mobile charges are distributed according to the Boltzmann distribution:
\begin{equation}\label{Eq37}
    \rho_{\mathrm{mobile}}=\sum\limits_i q_i C_i \exp{-\left (q_i\phi/k_\mathrm{B}T\right)}
\end{equation}
where $q_i$ and $C_i$ are the charge and the concentration of the $i^{th}$ type of mobile carrier, respectively. Following Debye and H\"{u}ckel, we linearize the exponent in Eq. \eqref{Eq37} and use the electroneutrality condition $\sum_i q_iC_i = 0$ to obtain:
\begin{equation}\label{Eq38}
\rho_{\mathrm{mobile}}=-\varepsilon\phi/L^2_\mathrm{D}
\end{equation}
where the Dybye length is defined as
\begin{equation}\label{Eq39}
    L^2_\mathrm{D}=\frac{\varepsilon k_\mathrm{B}T}{\sum_iq^2_iC_i}.
\end{equation}
Substituting Eq. \eqref{Eq38} into the Poisson Eq. \eqref{Eq4}, one obtains what is known as the Debye-H\"{u}ckel equation \cite{ref24} (or the linearized Poisson-Boltzmann equation). The generalization of the Debye-H\"{u}ckel equation for a point charge in quadrupolarizable media reads as
\begin{equation}\label{Eq40}
  \frac{1}{r^2}\frac{d}{dr}r^2\frac{d}{dr}\phi-\frac{L^2_Q}{r^2}\frac{d}{dr}r^2\frac{d}{dr}\frac{1}{r^2}\frac{d}{dr}r^2\frac{d}{dr}\phi=-\frac{q_i \delta(r)}{\varepsilon}+\frac{\phi}{L^2_\mathrm{D}}.
\end{equation}
We impose two boundary conditions to this equation. The first one is the standard electroneutrality condition$\int \rho \mathrm{d}V=0$. The second one is less orthodox - we require that the potential does not diverge at the origin, $\phi(0) < \infty$. The validity of the second condition is discussed in the previous Sec. \ref{Sec. 4.1}.The non-divergent at $r \to \infty$ solution of this equation is given by
\begin{equation}\label{Eq41}
  \phi=\frac{q_i}{4\pi\varepsilon}\frac{l^2_\mathrm{D}+l^2_Q}{l^2_\mathrm{D}-l^2_Q}\frac{\mathrm{e}^{-r/l_\mathrm{D}}-\mathrm{e}^{-r/l_Q}}{r}
\end{equation}
where we have introduced (as common for biharmonic equations) the two characteristic lengths $l_\mathrm{D}$ and $l_Q$ wich are related to $L_\mathrm{D}$ and $L_Q$ as:
\begin{equation}\label{Eq42}
 l_\mathrm{D}=L_Q\left (\frac{1}{2}-\frac{1}{2}\sqrt{1-4\frac{L^2_Q}{L^2_\mathrm{D}}}\right )^{-1/2},\quad \ l_Q=L_Q\left (\frac{1}{2}+\frac{1}{2}\sqrt{1-4\frac{L^2_Q}{L^2_\mathrm{D}}}\right )^{-1/2}.
\end{equation}
The inverse relations which define $L_\mathrm{D}$ and $L_Q$ through $l_\mathrm{D}$ and $l_Q$ are simpler
\begin{equation}\label{Eq43}
   L^2_\mathrm{D}=l^2_\mathrm{D}+l^2_Q,\qquad \ L^{-2}_Q=l^{-2}_\mathrm{D}+l^{-2}_Q.
\end{equation}
The potential in Eq. \eqref{Eq41} is finite and its value at $r = 0$ is
\begin{equation}\label{Eq44}
  \phi_0=\frac{q_i}{4\pi\varepsilon L_Q} \left (1+2 \frac{L_Q}{L_\mathrm{D}} \right )^{-1/2}.
\end{equation}
As was the case of a point charge in an insulator, the energy of the point charge is finite
\begin{equation}\label{Eq45}
    u_{\mathrm{el}}=\frac{q_i\phi_0}{2}=\frac{q^2_i}{8\pi\varepsilon L_Q}\left (1+2 \frac{L_Q}{L_\mathrm{D}} \right )^{-1/2}.
\end{equation}
In the limit $L_\mathrm{D} \to \infty$ Eq. \eqref{Eq41} simplifies to the potential of a point charge in an insulator (Eq. \eqref{Eq33}). In the case of negligible quadrupolarizability ($L_Q  \to 0$), Eq. \eqref{Eq41} reduces to the classical Debye-H\"{u}ckel potential of a point charge:
\begin{equation}\label{Eq46}
  \phi=\frac{q_i}{4\pi\varepsilon}\frac{\mathrm{e}^{-r/L_\mathrm{D}}}{r}
\end{equation}
The dependence Eq. \eqref{Eq42} of the characteristic lengths $l_\mathrm{D}$ and $l_Q$ on $L_\mathrm{D}/L_Q$ is analyzed in Fig. \ref{fig:2}. In dilute solutions, where $L_\mathrm{D} \gg 2L_Q$, both lengths $l_\mathrm{D}$ and $l_Q$ are real, and $l_\mathrm{D}$  is almost equal to $L_\mathrm{D}$ while $l_Q$ is almost equal to $L_Q$ (Fig. \ref{fig:2}, to the right of the bifurcation), which is the reason for the choice of indices. At a certain critical value of the Debye length ($L_\mathrm{D} = 2L_Q$), the lengths $l_\mathrm{D}$ and $l_Q$ become equal to each other. The critical concentration is $C_{\mathrm{cr}} =  \varepsilon k_\mathrm{B}T/8e^2L^2_Q$. At the critical value of $L_\mathrm{D}$ the potential in Eq. \eqref{Eq41} degenerates to the following result:
\begin{equation}\label{Eq47}
   \phi=\frac{q_i}{4\pi\varepsilon}\frac{\sqrt{2}}{L_\mathrm{D}}\mathrm{e}^{-\sqrt{2}r/L_\mathrm{D}}.
\end{equation}
This change in the functional dependence from $\exp({-r})/r$ to $\exp({-r})$ corresponds to a kind of "resonance" between the diffuse atmosphere of the mobile charges and the quadrupole moment cloud around a charge. When $L_\mathrm{D} < 2L_Q$, the two characteristic lengths in Eq. \eqref{Eq42} become complex and complex conjugate to each other (Fig. \ref{fig:2}, to the left of the bifurcation), i.e., the potential (Eq. \eqref{Eq41}) while diminishing with distance exhibits an oscillatory behavior. In this case one can rearrange Eq. \eqref{Eq41} in the form:
\begin{equation}\label{Eq48}
\phi=\frac{q_i}{4\pi\varepsilon} \frac{l^2_{\mathrm{Re}}-l^2_{\mathrm{Im}}}{l_{\mathrm{Re}}l_{\mathrm{Im}}r}\exp{\left (-\frac{l_{\mathrm{Re}}}{l^2_{\mathrm{Re}}+l^2_{\mathrm{Im}}}r\right )}\sin{\left (\frac{l_{\mathrm{Im}}}{l^2_{\mathrm{Re}}+l^2_{\mathrm{Im}}}r\right )}
\end{equation}
where $l_{\mathrm{Re}} = \textrm{Re} l_\mathrm{D}$ and $l_{\mathrm{Im}} = \textrm{Im} l_\mathrm{D}$. One can easily derive the following expressions for $l_{\mathrm{Re}}$ and $l_{\mathrm{Im}}$
\begin{equation}\label{Eq49}
    l_{\mathrm{Re}/\mathrm{Im}}=\frac{L_\mathrm{D}}{2}\sqrt{\pm1+ 2\frac{L_Q}{L_\mathrm{D}}}.
\end{equation}
It is well-known that oscillations of the electrostatic potential and the charge density exist \cite{ref25,ref26,ref27,ref28}; oscillation of wavelength $(l^2_{\mathrm{Re}} + l^2_{\mathrm{Im}})/l_{\mathrm{Im}}$ related to quadrupolarizability is, however, a fundamentally new phenomenon.

Let us now discuss the structure of the diffuse layer around a point charge in quadrupolarizable medium in relation with the classical Debye-H\"{u}ckel model. Physically, the quadrupolarizable medium does not support high gradients of the field, and therefore, non-zero quadrupolar length $L_Q$ results in a smoother potential and slightly expanded diffuse atmosphere. As a measure of this effect one can use the average distance between the central point charge and the diffuse charge of the double layer:
\begin{equation}\label{Eq50}
  L_{\mathrm{diffuse \, layer}}=\frac{\int r\rho \mathrm{d}V}{\int \rho \mathrm{d}V}=2\frac{L^2_\mathrm{D}+L_\mathrm{D}L_Q}{\sqrt{L^2_\mathrm{D}+2L_\mathrm{D}L_Q}}
\end{equation}
where the expressions for $\rho$   and  $\phi$ are taken from  Eqs. \eqref{Eq37} and \eqref{Eq41}, respectively. In the case of a low concentration of charges, Eq. \eqref{Eq50} can be expanded into series with respect to large $L_\mathrm{D}$ with the result $L_{\mathrm{diffuse \, layer}} \approx 2L_\mathrm{D} + L^2_Q/L_\mathrm{D} +...$. Thus, in the limit of the classical Debye-H\"{u}ckel model, the charge of the diffuse cloud stands at an average distance of $2L_\mathrm{D}$ from the central point charge. The quadrupolarizability of the medium leads to an expansion of the cloud to $2L_\mathrm{D} + L^2_Q/L_\mathrm{D}$.

\section[]{Electric field of a charged surface, a condenser and a dipolar surface in quadrupolar medium}\label{Sec. 5}

\subsection[]{A charged surface and a condenser}\label{Sec. 5.1}

Consider a surface of surface charge  $\rho^S$ (and zero dipole moment, $P^\mathrm{S}_z  = 0$) in a homogeneous insulator of dielectric permittivity $\varepsilon$  and quadrupolar length $L_Q$.  We solve the problem as if the surface is a field source immersed in a single medium, but it can be viewed as an interface between two dielectrics of equal $\varepsilon$ and $L_Q$ as well \cite{ref22}.

\begin{figure}[h]
\includegraphics[width=155pt]{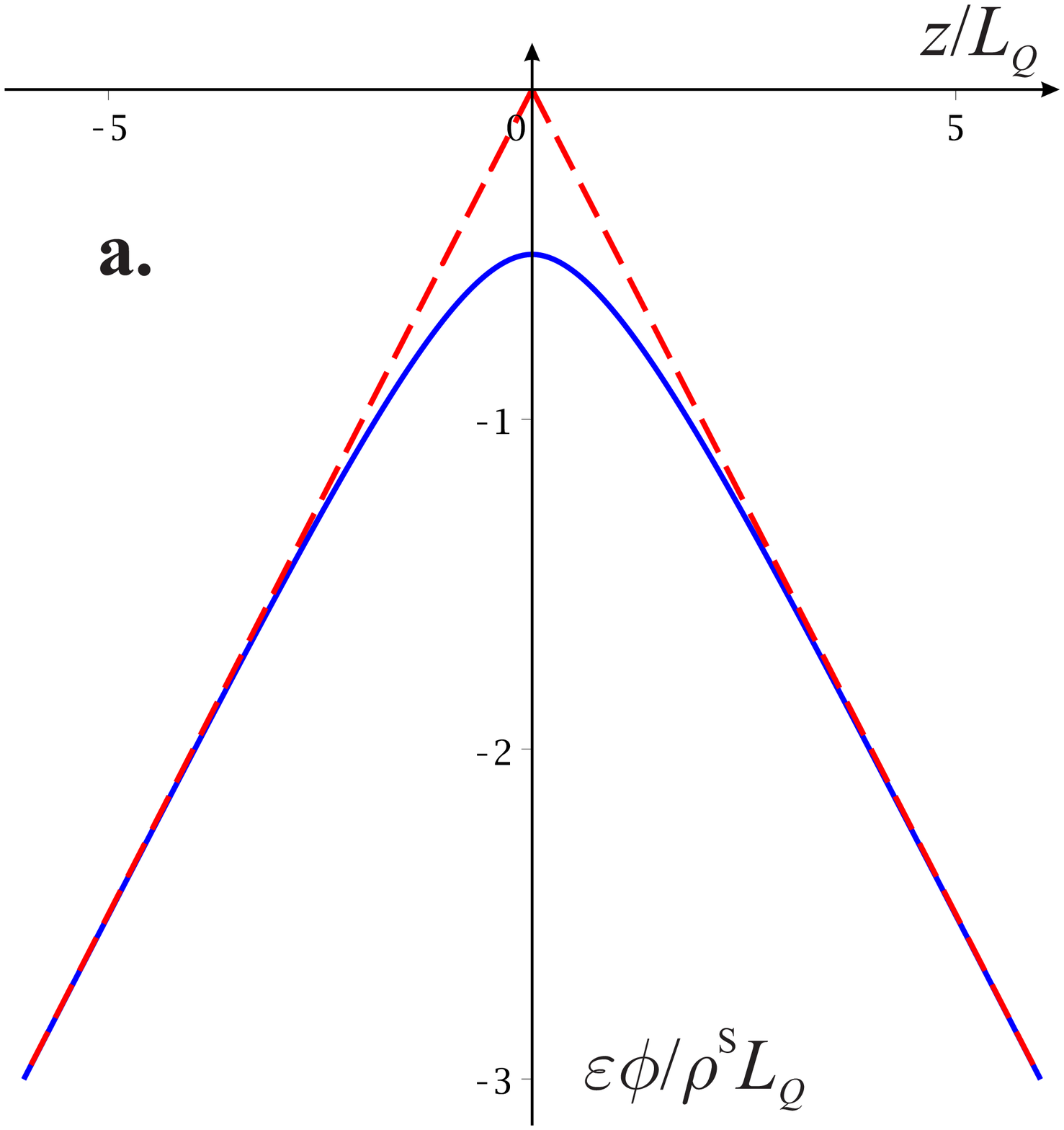}
\includegraphics[width=155pt]{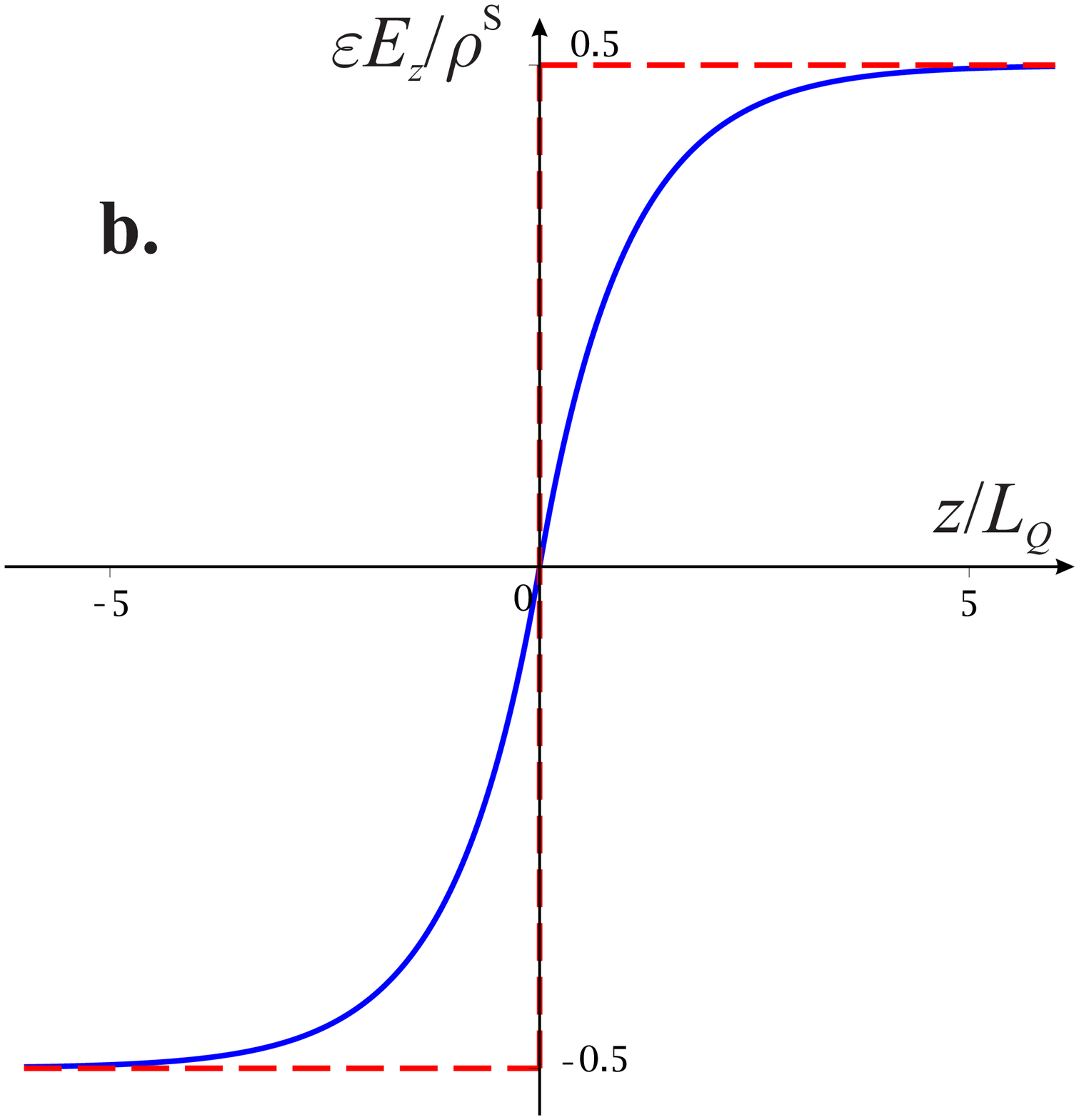}
\caption{Profile of \textbf{a}) the potential $\phi(z)$ and \textbf{b}) the field intensity $E_z(z)$ near a surface of charge $\rho^\mathrm{S}$ (suitably nondimensionalized). Comparison of the classical solution (red dashed line) with Eqs. \eqref{Eq54}-\eqref{Eq53}, following from the Coulomb's quadrupolar law.}\label{fig:3}
\end{figure}

The quadrupolar electrostatic equations Eq. \eqref{Eq8} for $E_z$ in the domain $z > 0$ (the field there is denoted by $E^+_{z}$) and $z < 0$ ($E^-_{z}$) read as:
\begin{equation}\label{Eq51}
    \frac{\mathrm{d} E^{\pm}_{z}}{\mathrm{d} z}-L^2_Q\frac{\mathrm{d} ^3E^{\pm}_{z}}{\mathrm{d} z^3}=0.
\end{equation}
The solution to this equation has to fulfil the Gauss law Eq. \eqref{Eq29} and the Graham-Raab boundary condition Eq. \eqref{Eq28}, not be diverging at infinity and to be an odd function, $E^+_{z}(z)=-E^-_{z}(-z)$. The general solution that fulfils these conditions is:
\begin{equation}\label{Eq52}
    E^{\pm}_{z}=\pm \left[1-A\mathrm{e}^{-|z|/L_Q}\right]\frac{\rho^\mathrm{S}}{2\varepsilon}
\end{equation}
The integration constant $A$ is determined employing the boundary condition for continuous field Eq. \eqref{Eq30}, namely, we require $E^+_{z}(0)=E^-_{z}(0)$, and we obtain that $A = 1$. Thus, the field of a charged surface turns out to be:
\begin{equation}\label{Eq53}
 E_{z}=\textrm{sg}(z)\frac{\rho^\mathrm{S}}{2\varepsilon}\left[1-\mathrm{e}^{-|z|/L_Q}\right]
\end{equation}
where $\textrm{sg}(z)$ is the signum function. The potential of a charged surface in a quadrupolar medium is obtained upon integration of $-E_z$ with respect to $z$
\begin{equation}\label{Eq54}
    \phi=-\frac{\rho^\mathrm{S}}{2\varepsilon}\left(|z|+L_Q \mathrm{e}^{-|z|/L_Q}\right ).
\end{equation}
It is easy to see that the derivative of $E_z$ has finite value at $z =0$:$\mathrm{d} E_{z}/\mathrm{d}z|_{z=0}= \rho^\mathrm{S}/2\varepsilon L_Q$. In Fig. \ref{fig:3} we compare these results with the corresponding ones from the classical dipolar theory $E_{z}=\textrm{sg}(z)\rho^\mathrm{S}/2\varepsilon$ and $\phi=-\rho^\mathrm{S}|z|/2\varepsilon$.

Let us consider now two charged surfaces located at $z = h/2$ with surface charge density $\rho^\mathrm{S}$ and at $z = -h/2$ with charge density $-\rho^\mathrm{S}$, respectively. This is the problem for a condenser of finite thickness $h$ immersed in a quadrupolar medium. The easiest way to obtain the respective field is to use Eq. \eqref{Eq52} as a Green's function for the problem. The total field intensity in the three domains denoted by superscripts "+", "i" and "-" corresponding to $z < -h/2$, $h/2 > z > -h/2$ and $z>h/2$, respectively, is obtained by adding the fields of the two charged surfaces with the result:
\begin{eqnarray}\label{Eq55}
                E^+_{z} & = & -\left (e^{-|z-h/2|/L_Q}-e^{-|z+h/2|/L_Q}\right )\frac{\rho^\mathrm{S}}{2\varepsilon} \nonumber \\
                E^\mathrm{i}_{z} &=& \left (e^{-|z-h/2|/L_Q}+e^{-|z+h/2|/L_Q}\right )\frac{\rho^\mathrm{S}}{2\varepsilon}- \frac{\rho^\mathrm{S}}{\varepsilon} \\
                E^-_{z} &=& \left (e^{-|z-h/2|/L_Q}-e^{-|z+h/2|/L_Q}\right )\frac{\rho^\mathrm{S}}{2\varepsilon}. \nonumber
              \end{eqnarray}
The limit as $h \to 0$ of the piecewise function Eq. \eqref{Eq55} corresponds to the case of an infinitely thin condenser:
\begin{equation}\label{Eq56}
  E^\pm_{z}=-\frac{\rho^\mathrm{S}}{2\varepsilon}\frac{h}{L_Q}\mathrm{e}^{-|z|/L_Q}.
\end{equation}
More importantly, this limit allows us to consider the problem of a surface with surface density of the dipolar moment $P^\mathrm{S}_z=h\rho^\mathrm{S}$.

\subsection[]{A dipole moment-carrying surface (infinitely thin condenser)}\label{Sec. 5.2}

\begin{figure}[h]
\includegraphics[width=155pt]{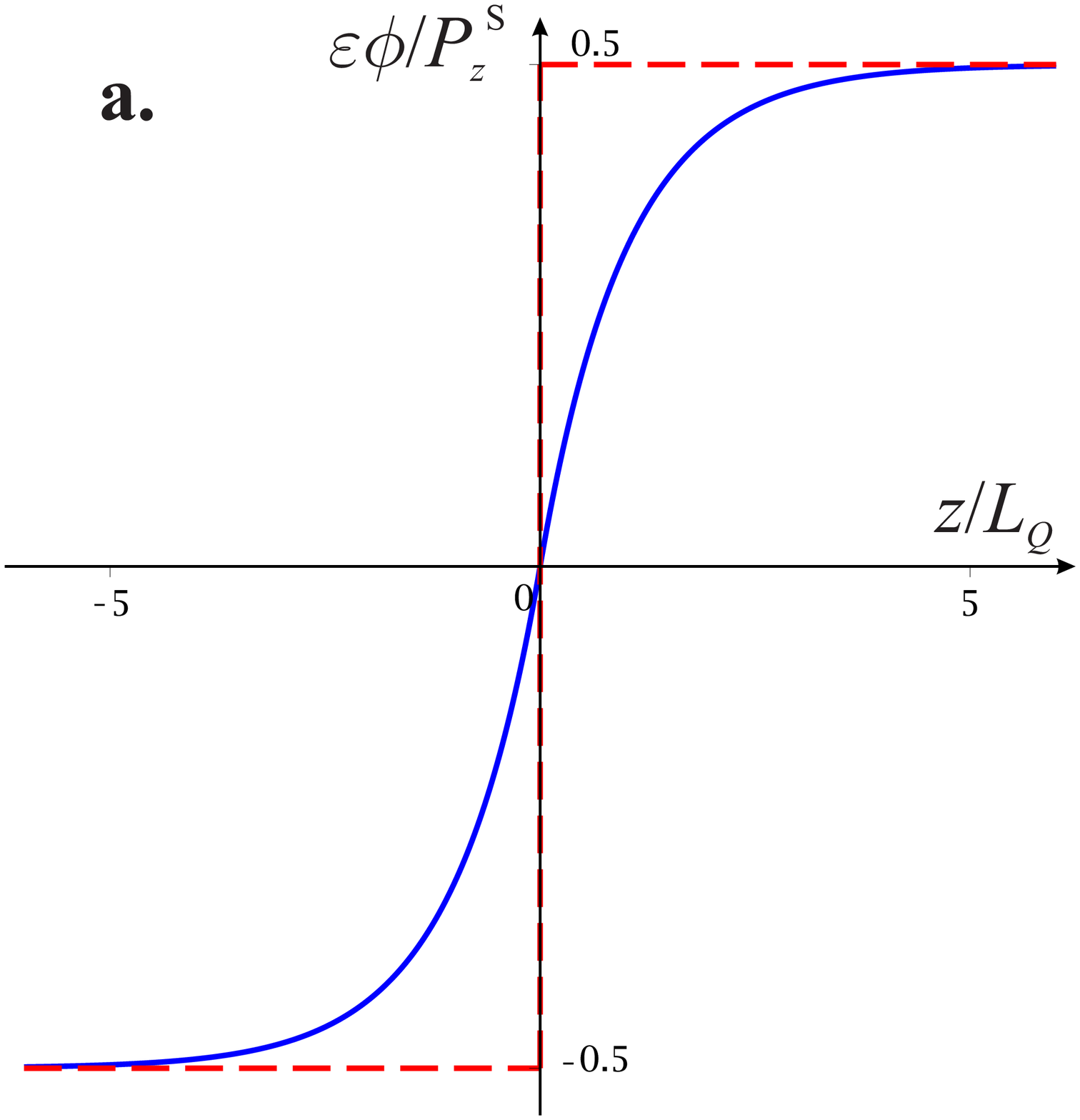}
\includegraphics[width=155pt]{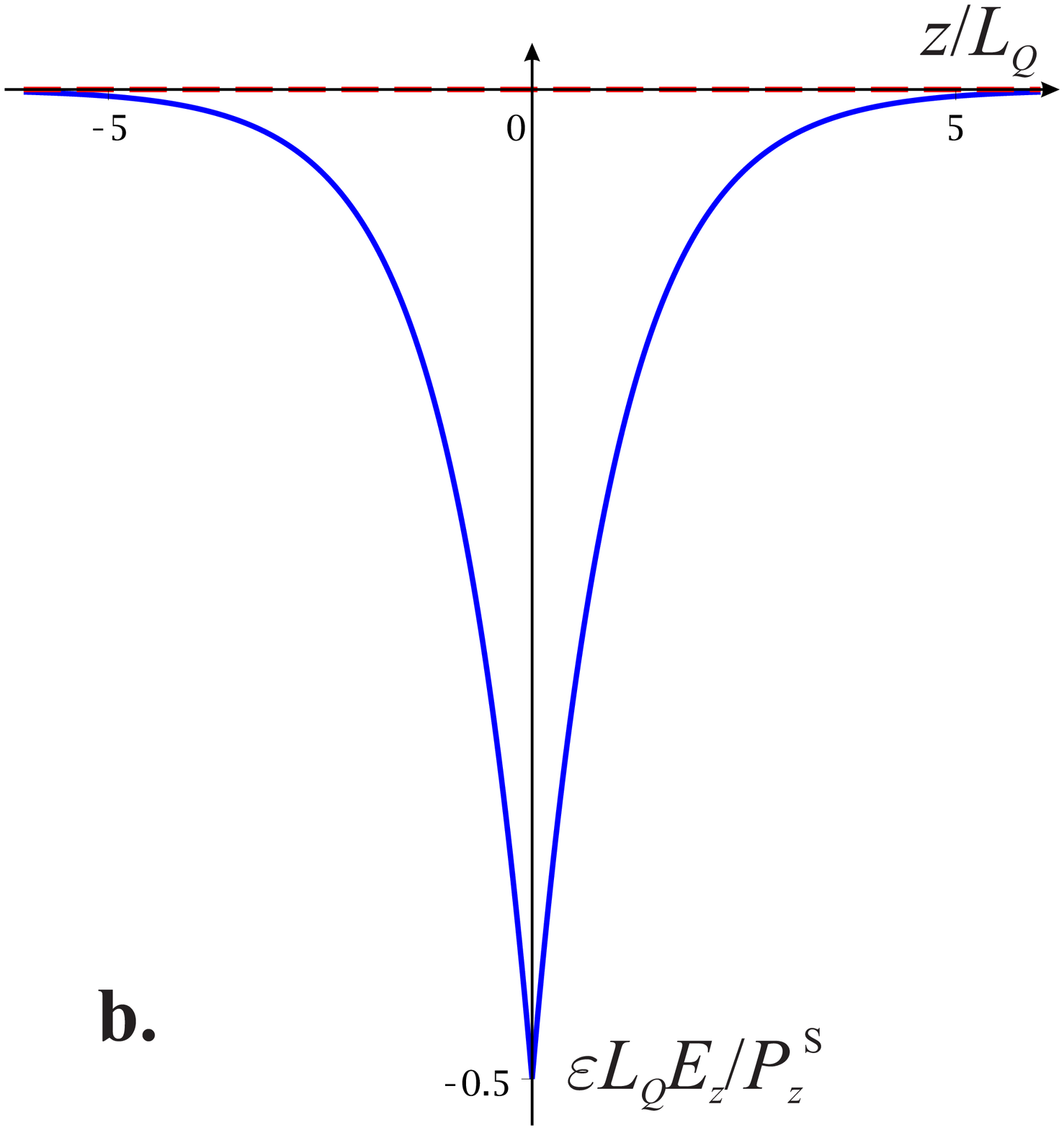}
\caption{Profile of \textbf{a}) the electrostatic potential $\phi(z)$ and \textbf{b}) field intensity $E_z(z)$ created by an infinitely thin capacitor of surface dipole moment $P^\mathrm{S}_z$ located in $z = 0 $ ($\phi$, $E_z$ and $z$ are suitably  nondimensionalized). The figure compares the classical solution of Poisson's equation (red dashed line) with Eqs. \eqref{Eq59}-\eqref{Eq60} that follow from the quadrupolar equations of Maxwell.}\label{fig:4}
\end{figure}

Consider a surface of dipole moment $P^\mathrm{S}_z$ and zero surface charge density in a homogeneous insulator with quadrupolar length $L_Q$ and dielectric constant $\varepsilon$ \cite{ref22}. This can serve as a model for a lipid bilayer and for certain defective structures in crystals. The solution of Eq. \eqref{Eq51} that does not diverge at infinity is:
\begin{equation}\label{Eq57}
    E^{\pm}_{z}=A^{\pm}\mathrm{e}^{-|z|/L_Q}.
\end{equation}
where $A^+$ and $A^-$ are integration constants. This solution fulfils Gauss's law Eq. \eqref{Eq29} for any value of the integration constants $A^+$ and $A^-$. The Graham-Raab multipole condition Eq. \eqref{Eq28} gives the relation:
\begin{equation}\label{Eq58}
  A^++A^-=-\frac{P^\mathrm{S}_z}{\varepsilon L_Q}.
\end{equation}
In order to determine the second constant, we invoke the symmetry of the problem, namely, the potential of the system must be an odd function and $E_z$ must be even function of $z$, i.e., $E^+_z(z) = E^-_z(-z)$.  The constants are determined as $A^+ = A^- = -P^\mathrm{S}_z/2\varepsilon L_Q$. Upon substituting in Eqs. \eqref{Eq57} and \eqref{Eq58} the final solution for the field is obtained in the form:
\begin{equation}\label{Eq59}
  E_{z}=-\frac{P^\mathrm{S}_z}{2\varepsilon L_Q}\mathrm{e}^{-|z|/L_Q}.
\end{equation}
As could be expected, this result coincides with the expression for the field of an infinitely thin condenser (recall that $P^\mathrm{S}_z=h\rho^\mathrm{S}$. The integration of Eq. \eqref{Eq59} gives the potential:
\begin{equation}\label{Eq60}
    \phi=\textrm{sg}(z)\frac{P^\mathrm{S}_z}{2\varepsilon}\left(1-\mathrm{e}^{-|z|/L_Q}\right ).
\end{equation}
The potential difference between $z \to \infty$ and $z \to - \infty$ is:
\begin{equation}\label{Eq61}
    \phi(z \to \infty)-\phi(z \to -\infty)=\frac{P^\mathrm{S}_z}{\varepsilon}.
\end{equation}
Recall that the same potential difference is obtained in the dipolar electrostatics as well (cf. e.g. {\it Sec. 14} of Ref. \cite{ref29}). The solution in Eqs. \eqref{Eq59} and \eqref{Eq60} is presented in Fig. \ref{fig:4} together with a comparison with the corresponding results from the dipolar electrostatics $\phi=\textrm{sg}(z)P^\mathrm{S}_z/2\varepsilon$ and $E_{z}=-\delta(z) P^\mathrm{S}_z/\varepsilon$.

We will conclude this subsection with three final remarks. First, within the quadrupolar electrostatic laws and Eq. \eqref{Eq55}, a capacitor creates field that penetrates outside the plates of the condenser. This phenomenon has no analogue in the frame of Poisson's electrostatics. Second, the comparison between the classical and the quadrupolar solution of the electrostatic problems illustrated in Fig. \ref{fig:3} and Fig. \ref{fig:4} demonstrates two features of quadrupolar electrostatics: first, the regularization of the solution for the field (a charged surface creates continuous $E_z$ and $\mathrm{d}E_z/\mathrm{d} z$ and only the second derivative is discontinuous - compare to the classical discontinuity of $E_z$; a dipolar surface creates continuous field and only $\mathrm{d}E_z/\mathrm{d} z$ has a discontinuity - compare to the classical discontinuity of $\phi$). This regularization was already observed with the point charge problem in Ref. \cite{ref09} (Section \ref{Sec. 4.1}). The last remark is that the field is continuous only if on both sides of the surface there exist quadrupolar media. At the boundary between quadrupolar and non-quadrupolar medium, the boundary condition for $\textbf{E}$ (Eq. \eqref{Eq30}) does not hold. In this case, however, no fourth boundary condition is required.

\section[]{Conclusion}
The present work summarizes the main results of our previous studies \cite{ref09,ref21,ref22,ref23}. We investigate the effect of taking into account the presence of quadrupoles in the continuous medium. For this purpose we derive a new equation of state for the quadrupolarization $\textbf{Q}$ (Eq. \eqref{Eq7}) and generalize the classic Poisson's equation (Eq. \eqref{Eq8}) and the required boundary conditions (Eqs. \eqref{Eq27}-\eqref{Eq30}) for quadrupolar medium. When we apply these equations to some basic electrostatic problems we obtain results which have no analogue within the classical electrostatics:
i)	the potential of a point charge in quadrupolar medium and its self-energy are finite even at the position of the charge;
ii)	the potential of a point charge in conducting media has oscillatory behaviour above certain critical concentration of the charges;
iii)	the electric field of a charged surface is a continuous function at the surface;
iv)	the electric field penetrates outside the plates of a condenser placed in quadrupolar medium.
Therefore, the following conclusions for the characteristic features of the quadrupolarizable media can be drawn: taking into account the presence of quadrupoles in the media makes the potential smoother and damps the electric field gradients.

\section*{Acknowledgements}
The work is funded by National Science Fund through Contract 51 from 12.04.2016 with Sofia University.

\section*{Citation \& References}

\end{document}